\documentclass[prl,aps,twocolumn,showpacs]{revtex4}
\include{graphics}
\usepackage{epsfig}
\usepackage{graphicx}
\usepackage{textcomp}

\begin{document}
\title{Experimental Study of Parametric Autoresonance in Faraday Waves}
\author{Oded Ben-David, Michael Assaf, Jay Fineberg, and Baruch Meerson}
\affiliation{The Racah Institute of Physics, The Hebrew University
of Jerusalem, Jerusalem 91904, Israel}
\begin{abstract}
The excitation of large amplitude nonlinear waves is achieved via
parametric autoresonance of Faraday waves. We experimentally
demonstrate that phase locking to low amplitude driving can
generate persistent high-amplitude growth of nonlinear waves in a
dissipative system. The experiments presented are in excellent
agreement with theory.
\end{abstract}
 \maketitle

\textit{Introduction.} When a nonlinear oscillator is resonantly
driven by small amplitude periodic forcing, the amplitude growth is
arrested, even at zero dissipation, when nonlinearity comes into
play. This is because a frequency mismatch develops between the
(constant) driving frequency and the (amplitude dependent)
oscillator frequency \cite{Bogoliubov}. \textit{Persistent}
amplitude growth can be achieved, by autoresonance, when the system
nonlinearly locks to an externally varied (``chirped") driving
frequency to retain resonant conditions. The precise form of the
chirp is unimportant once its sign is correct, and the chirp rate is
below a critical value. First predicted for harmonic 
forcing, autoresonance has found many applications
\cite{Friedland1}. The technique was extended to weakly dissipative
oscillators \cite{Fajans_damping} and to nonlinear waves and
vortices in non-dissipative systems \cite{Meerson5}. The theory of
\textit{parametric} autoresonance (PAR) was recently developed,
first for nonlinear oscillators \cite{Khain} and later \cite{assaf}
for nonlinear Faraday waves: standing gravity waves on a free
surface of a fluid which are excited parametrically by vertical
vibrations. This theory \cite{assaf} predicts that a downward chirp
of the vibration frequency should cause persistent wave growth,
which is only expected to terminate at large amplitudes, when an
underlying \textit{constant} frequency system (CFS), introduced
below, ceases to exhibit a non-trivial stable fixed point.

Here we report the first experimental verification of PAR excitation
of a nonlinear wave. Using Faraday waves we demonstrate that
autoresonance is not hindered by moderate dissipation, and the
results compare well with an extended version of the theory
\cite{assaf}. We show that the predicted (negative) frequency chirp
indeed drives persistent wave growth, via the PAR mechanism, to
amplitudes that surpass the theory's region of validity.

\textit{Theory.} The theory of PAR excitation of nonlinear Faraday
waves \cite{assaf} is based on an amplitude expansion that extends
earlier constant-frequency treatments \cite{Miles,douady} to the
chirped frequency case. Here we summarize the main predictions of
Ref. \cite{assaf} and extend the model by (i) introducing a more
general form of driving acceleration, and (ii) taking a more
complete account of dissipation.

Throughout this Letter we consider a rectangular fluid cell of
length $l$, width $w$  and depth $h$ in the $x$, $y$, and $z$
direction, respectively. To avoid three-wave interactions
\cite{vinals} we assume that the surface tension corrections are
small \cite{pure}. Furthermore, we assume a deep water limit $h>l$
and a sufficiently small $w$ so that the fluid motion is
two-dimensional, depending on $x$, $z$ and $t$. The vertical
displacement of the vibrating cell is $\zeta(t)=\zeta_0 (t)\cos
\Phi(t)$, where $\Phi (t)=\int_0^t\omega(t^{\prime}) \, dt^{\prime}$
is the driving phase, while the driving frequency $\omega(t)$ and
amplitude $\zeta_0(t)$ vary slowly on the time scale of
$\omega^{-1}$. In the weakly nonlinear regime, the (time-dependent)
scaled acceleration of the cell is
$\varepsilon(t)=\omega^{2}(t)\zeta_{0}(t)/g\ll\,1$, where $g$ is the
gravity acceleration. As a result, the wave steepness parameter $k
\eta_{max} \ll {1}$, where $k=\pi/l$ is the wave number of the
fundamental mode, and $\eta$ is the wave amplitude. As the nonlinear
frequency shift of standing gravity waves, in the deep-water limit,
is negative \cite{Miles}, the PAR driving must use a negative
frequency chirp: $d\omega/dt<0$. The amplitude dynamics of the
fundamental mode are governed, at leading nonlinear order, by the
equation \cite{assaf}:
\begin{equation}\label{motionchirp}
\ddot{\eta}_{1}\!+2\!\gamma\dot{\eta}_{1}\!+
\frac{k^{2}}{2}(5\dot{\eta}_{1}^{2}\eta_{1}\!-
\!3\Omega^{2}\eta_{1}^{3})\!+\Omega^{2}\!
\left[\!1\!+\!\varepsilon(t)\cos\Phi(t)\right]\!\eta_{1}\!=\!0{,}
\end{equation}
where $\Omega =(kg)^{1/2}$ and $\gamma\ll\Omega$ are the linear wave
frequency and \textit{effective} linear damping rate, respectively
(see Ref. \cite{christiansen} for a review of different
contributions to $\gamma$). Higher order modes are enslaved to
$\eta_{1}$ and can be calculated once $\eta_1$ is found. For
concreteness, we assume a linear chirp:
$\omega(t)=\omega_{0}-\mu\,t$, where $\mu
> 0$ is constant,  and $\omega_{0} = 2 \Omega$ is the
resonant value of the driving frequency. We also assume
$\varepsilon(t)=\varepsilon_{0}(1+\beta\,t)$, where
$\varepsilon_{0}>0$ and $\beta>0$ are constant \cite{eps}. Now we
make an ansatz
$\eta_{1}(t)=A(t)\cos\left[\Omega{t}+\varphi(t)\right]$, where $A$
and $\varphi$ are the slowly varying amplitude and phase, and use
the averaging method \cite{Bogoliubov}. Rescaling time
$\tau=(\varepsilon_{0}\Omega{t})/4$, amplitude
$B=kA\varepsilon_{0}^{-1/2}$, chirp rate
$m=8\mu/(\varepsilon_{0}\Omega)^{2}$ and damping rate
$\Gamma=4\gamma/(\varepsilon_{0}\Omega)$, and denoting $\phi(t)=\mu
t^2/2+\varphi(t)$ and
$\tilde{\beta}=4\beta/(\varepsilon_{0}\Omega)$, we obtain:
\begin{eqnarray}
\dot{B}&=&(1+\tilde{\beta}\tau)B\sin(2\phi)-\Gamma\,\!B\,,\nonumber\\
\dot{\phi}&=&(1+\tilde{\beta}\tau)\cos(2\phi)-B^{2}+m\tau\,,
\label{averaged}
\end{eqnarray}
where the dots stand for derivatives with respect to the slow time
$\tau$. The underlying CFS corresponds to $m=\tilde{\beta}=0$. Let
us start the frequency chirp from the steady state obtained for a
constant-frequency driving (which is the stable fixed point of the
underlying CFS). For small $m$ and $\tilde{\beta}$, the PAR wave
growth corresponds to the stable \textit{quasi}-fixed point of
Eqs. (\ref{averaged}). To leading order
\begin{eqnarray}
B_{*}^2(\tau)&=&\left[(1+\tilde{\beta}\tau)^{2}-\Gamma^{2}\right]^{1/2}+m\tau\,,\nonumber\\
\phi_{*}(\tau)&=&\frac{1}{2}\arcsin\left(\frac{\Gamma}{1+\tilde{\beta}\tau}\right)\,.
\label{trends}
\end{eqnarray}
The PAR breaks down if the rescaled chirp rate $m$ exceeds a
critical value $m_{cr}=\mathcal{O}(1)$ which depends on $\Gamma$
\cite{assaf} and $\tilde{\beta}$.  In any case, the wave growth must
terminate at large amplitudes, when higher-order corrections to Eqs.
(\ref{motionchirp}) and (\ref{averaged}) cause the disappearance of
the non-trivial stable fixed point in the underlying CFS
\cite{Miles,douady}.

Alternatively, we can start from a very small initial amplitude
$B_0$ far from resonance and pass through the resonance. In the
\textit{linear locking stage}, we can drop the $B^2$ term in Eq.
(\ref{averaged}) and obtain \cite{assaf}:
\begin{eqnarray}
B_{*}^2(\tau)&\simeq&B_{0}^2\exp\!\left[\tau\sqrt{1\!-\!(m\tau)^{2}}
+\frac{\arcsin(m\tau)}{m}-\!2\Gamma\tau \right]\,,\nonumber\\
\phi_{*}(\tau)&\simeq&\frac{\pi}{4}+\frac{1}{2}\arcsin(m\tau)-\frac{m}{4\sqrt{1\!-\!(m\tau)^{2}}}\,,
\label{passage}
\end{eqnarray}
where we have put $\tilde{\beta}=0$. As long as $B_*(\tau) \ll 1$,
Eqs. (\ref{passage}) are valid on the time interval $-1<m\tau<1$
(but not too close to $m\tau=1$ \cite{assaf}). Remarkably, Eqs.
(\ref{passage}) correspond to an \textit{unstable} (saddle)
quasi-fixed point \cite{assaf}, so the system eventually escapes
from this point and either enters the \textit{nonlinear}
phase-locking regime, described by Eqs. (\ref{trends}) with a
shifted time, or loses phase-locking. If/when $B_*(\tau)$ approaches
unity, Eqs. (\ref{passage}) become invalid. The amplitude
$B_{*}(\tau)$, given by the first of Eqs. (\ref{passage}), reaches a
maximum at $\tau_{m}=\sqrt{1-\Gamma^{2}}/m$. Its maximum value
\begin{equation}\label{maxamp}
B_*^{max} \simeq\,B_{0}\exp\!\left[\!\frac{1}{2m}\!\left(\arccos
\Gamma- \Gamma\sqrt{1\!-\!\Gamma^{2}}\right)\!\right]
\end{equation}
decreases with an increase of $m$. Therefore, at sufficiently large
$m$ Eqs. (\ref{passage}) and (\ref{maxamp}) remain valid over the
whole interval $-1<m\tau<1$.

\textit{Experiment.} Our experiments were conducted in a transparent
cell mounted on a Unholtz-Dickie model 5PM electro-mechanical shaker
made to oscillate in the vertical ($z$) direction. At $\omega_0=54.7
\pm 0.16$ sec$^{-1}$ we excite an almost pure gravity wave
\cite{pure} with $k=2\pi/8$ $\mbox{cm}^{-1}$, whose wavelength is
twice the cell length of $l=4$ cm. The cell, of width $w=2$ cm, was
filled to depth $h \simeq 6$ cm with hexamethyldisiloxane, and
sealed to prevent evaporation. Hexamethyldisiloxane is a Newtonian
fluid whose kinematic viscosity, surface tension and density are,
respectively, $0.65$ cSt, $15.6$ dyne/cm and $0.76$ g/cc. The
kinematic viscosity
 was stabilized to within $\pm 1.5\%$ by fixing the fluid
temperature to $26.6 \pm 0.2$ \textcelsius. We generated
acceleration profiles of the form $a(t)\cos(\omega_0 t -
{\mu}t^2/2)$ by computer.  $a(t)$ was controlled to $1\%$ and
measured to $0.001\,g$ resolution by an ADXL103 accelerometer.

\begin{figure}[ht]
\includegraphics[width=0.95\columnwidth,clip=true,keepaspectratio=true]{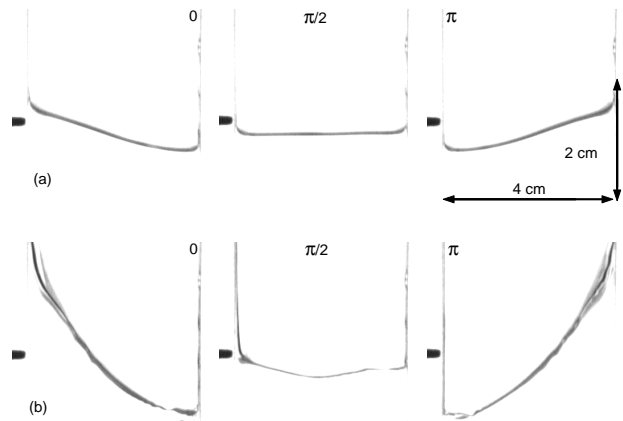}
\caption{Images at different phases of (a) (top) an initial wave
state at $\omega_0$ and $a=0.064$ g and (b) (bottom) the same state
after being autoresonantly driven at a chirp rate of $\mu=0.2$
sec$^{-2}$ to $\delta \equiv (\omega_{0}-\omega)/\omega_{0}=0.08$
and $a=0.113$ g.} \label{fig1}
\end{figure}

Computer-triggered visualization of the wave profile was performed
by uniform illumination of the fluid-air interface from behind.
The interface's high curvature, due to its wetting of the side
walls, refracted light away from a CCD camera mounted on the
opposite side of the cell. This resulted in a sharp dark edge
depicting the interface, see Fig. \ref{fig1}. A reference mark on
the cell enabled measurement of its instantaneous vertical
position. Edge detection produced a vector of the interface's
location relative to the cell position as a function of time.
The scaled amplitude of the fundamental, $k A$, measures the
steepness of the wave profile. To enable direct comparison to
theory, we needed to isolate the fundamental of the interface
waveform. To this end, we measured the difference in wave elevation
between two points, chosen to be symmetrical about the center of the
cell on the $x$-axis. This eliminates, by symmetry, all of the even
harmonics of the interface elevation. Although the third harmonic is
not filtered out, the resulting systematic error of $A$ is only
about $1 \%$ at $kA=0.2$, and does not exceed $6\%$ for $kA=0.6$. We
could therefore ignore the third harmonic while comparing our
measurements with the theory in the weakly nonlinear regime. The
statistical error in $k A$, as estimated from steady state data, is
$\sim1\%$. The instantaneous phase mismatch $\phi$ between the
driving and temporal response was extracted using complex
demodulation \cite{demodul} of two time series: of the measured
reference mark on the cell and of the wave elevation.

\begin{figure}[ht]
\includegraphics[width=0.95\columnwidth,clip=true,keepaspectratio=true]{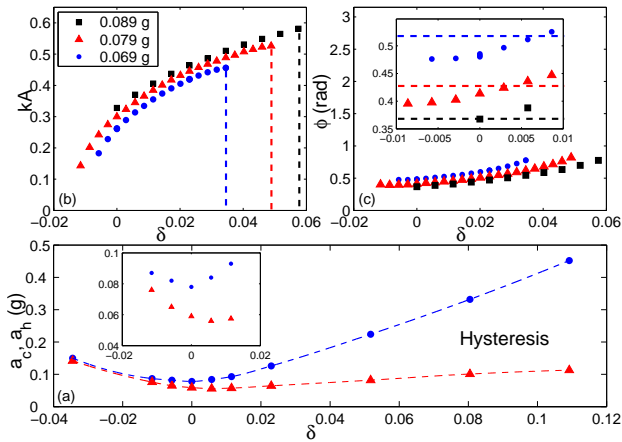}
\caption{(bottom) (a) Measured $\delta$-dependence of $a_c$
(circles) and the hysteretic region bounded by $a_h$ (triangles).
Dashed lines depict cubic interpolation. Inset: closeup of the
vicinity of $\delta=0$. (top) Steady-state measurements of ${k}{A}$
(b) and $\phi$ (c) for different values of $a=const$ (legend) versus
$\delta$. Here ${\Gamma}{\geq}0.65$. Dashed lines in (b) denote the
maximum detuning attainable for each value of $a$. Inset in (c) is a
closeup of the vicinity of $\delta=0$, dashed lines depict the
predicted values. } \label{fig2}
\end{figure}

In Fig. \ref{fig2}(a) we present the measured critical
accelerations $a_c(\delta)$ for the Faraday instability of the
flat surface as function of the scaled detuning $\delta\equiv
(\omega_{0}-\omega)/\omega_{0}$. The system undergoes a hysteretic
transition \cite{douady} at another critical acceleration,
$a_h(\delta)$: the lowest acceleration at which the nonlinear wave
remains stable.

Figures \ref{fig2}(b) and \ref{fig2}(c) depict steady-state (CFS)
measurements of $k A$ and $\phi$, respectively, as a function of
$\delta$ for $a=const$. Note that $a=a_h$ sets the maximum
attainable detuning. Until $a=a_h$, $k A$ increases rapidly with
increasing $\delta$ (decreasing $\omega$). Beyond this point, waves
will rapidly decay.

The 
wave damping rate $\gamma$ can be extracted from measurements of
$a_c$, since $\Gamma=1$ at the instability onset. With $\gamma$ in
hand, we can directly compare our measurements to theoretical
predictions with no other free parameters. The measured phase
difference for $\delta=0$ agrees within $5\%$ with the predicted
value. The slow increase in $\phi$ with $\delta$ in Fig.
\ref{fig2}(c) is due to higher order nonlinearities.

Our first series of measurements used a linear chirp,
$\delta=\mu{t}/\omega_{0}$, starting from a steady state wave with a
small but finite amplitude, at $\omega=\omega_0$ and $a=a_0
> a_h$. Importantly, our choice of the system parameters
precluded the excitation of other linear modes during a negative
chirp. As shown in Fig. \ref{fig2}(a), $a_h(\delta)$ is almost
linear with $\delta$ for $0\leq\delta<0.12$. To maximize the
frequency range of the excitation, we ramped $a$ linearly in time,
$a(\delta)=a_0+\alpha\delta$, to make $a$ nearly parallel
($\alpha=0.609$ g) to $a_h(\delta)$.

\begin{figure}[ht]
\includegraphics[width=0.95\columnwidth,clip=true,keepaspectratio=true]{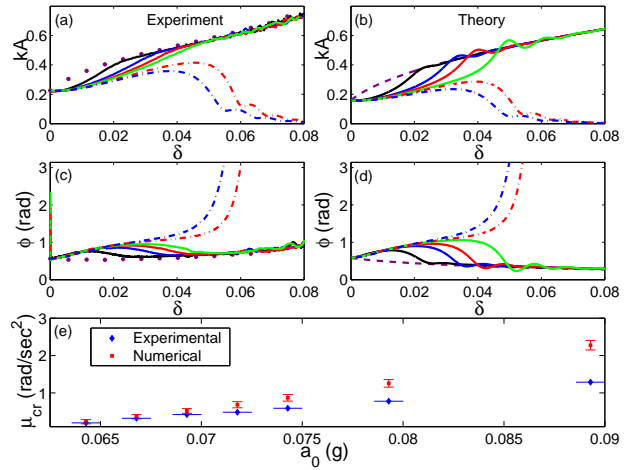}
\caption{Chirps in $\delta$ initiated from a fixed point. Measured
(a) and computed from Eq. (\ref{averaged}) (b) values of ${k}{A}$
versus the detuning $\delta$, starting from $\delta=0$ and
$a_0=0.064$g. $\Gamma\simeq{0.9}$. Chirp rates of $\mu=0.09$
(black), $0.15$ (blue), $0.18$ (red), and $0.20$ (green) sec$^{-2}$
converge to the fixed point (circles). ${k}{A}$ for $\mu=0.21$ and
$0.25$ sec$^{-2}$ (red and blue dash-dotted lines, respectively)
diverge from the steady state and decay. The dashed line in (b)
depicts the line of quasi-fixed points from Eq. (\ref{trends}). (c)
and (d): measured and computed values of $\phi$, respectively. (e)
Experimental (diamonds) and computed from  Eq. (\ref{averaged})
(squares) values of the critical chirp rate $\mu_{cr}$ as a function
of $a_0$.} \label{fig3}
\end{figure}

An example of an autoresonant state achieved in a typical ``ramping"
experiment is shown in Fig. \ref{fig1}, where images at the initial
and final values of $\delta$ are presented.  Note the substantial
increase in the wave amplitude. In Fig. \ref{fig3} (a)-(d) we
analyze the dynamics leading from Fig. \ref{fig1}(a) to Fig.
\ref{fig1}(b) by comparing experimental measurements of $kA$ and
$\phi$ with theoretical predictions, obtained by integrating Eqs.
(\ref{averaged}) numerically for several chirp rates for the same
ramp and initial conditions \cite{direct}. For small values of
$\mu$, phase-locking occurs where, as a function of $\delta$, both
$kA$ and $\phi$ quickly converge to the values that they would
attain in steady-state for each instantaneous value of $a(t)$ and
$\delta(t)$, see Eqs. (\ref{trends}). At larger values of $\mu$,
however, {\it no} phase locking occurs. Here $\phi$ diverges rapidly
away from the steady-state curves, and subsequently $kA$ rapidly
decays to zero.

As the theoretical curves were obtained with {\it no} free
parameters, the agreement between experiment and theory is striking.
Although the transients are slightly more long-lived in the theory,
identical convergence/divergence of both the phase and amplitude of
the waves to/from their steady-state values is observed for all of
the values of $\mu$ used.

An important prediction of the theory is that above an
$a_0$-dependent critical chirp rate, $\mu_{cr}$, phase-locking is
not possible. For relatively small values of $a_0$, there is good
quantitative agreement between the measured and predicted values of
$\mu_{cr}$, see Fig. \ref{fig3}(e). Furthermore, $\mu_{cr}$ still
exists for larger accelerations. Even though $\mu_{cr}$ increases
with $a_0$, as predicted, the predicted and observed values of
$\mu_{cr}$ systematically diverge with increasing $a_0$. This
divergence is not surprising, as for $a_0 > 0.07$ g the
phase-locking occurs for ${k}{A} > 0.4$, where we would expect the
weakly nonlinear theory to become inaccurate. These results imply
that PAR for $\mu<\mu_{cr}$ persists far beyond the region of
validity of the weakly nonlinear approximation. Note that transient
stages of $\phi$ in Fig. \ref{fig3} trace an envelope, corresponding
to the universal trajectory (a saddle point) obtained at
$\mu=\mu_{cr}$ \cite{assaf}.

Our second series of measurements involve ``passing through" the
resonance at $\omega_0$ before any waves are initially excited. We
begin from a flat state at a negative detuning $\delta_{init}<0$
(i.e. $\omega>\omega_0$), with $a_c(\delta=0)<a<a_c(\delta_{init})$.
We then apply a linear chirp $\delta=\mu{t}/\omega_{0}$ while
keeping $a$ fixed throughout the experiment. As $\delta$ is
increased, we pass through a region of $\delta$ where
$a>a_c(\delta)$.

\begin{figure}[ht]
\includegraphics[width=0.95\columnwidth,clip=true,keepaspectratio=true]{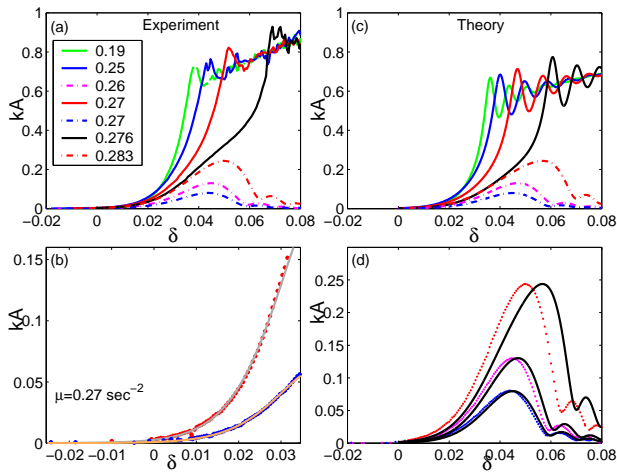}
\caption{Passing through resonance. (a) Measurements of ${k}{A}$ for
$a=0.168$ g, $-0.046 < \delta < 0.08$. Solid (dashed-dotted) lines
depict phase-locking (unlocking) for the values of $\mu$
(sec$^{-2}$) presented in the legend. (b) Comparison of measured
(dots) and computed from Eq. (\ref{passage}) (lines) values of
${k}{A}$ in the initial stages of the two runs with $\mu=0.27$
sec$^{-2}$. The different initial values of ${k}{A_0}$ [obtained by
fitting Eq. (\ref{passage})], resulting from low-level noise,
determine the eventual behavior. (c) Computed values of ${k}{A}$,
using Eq. (\ref{averaged}), corresponding to the parameters of (a).
(d) Comparison of computed (solid lines) and measured (dotted lines)
values of ${k}{A}$  for ${k}{A} <  0.3$.} \label{fig4}
\end{figure}

The dynamics of passing through resonance are demonstrated in Fig.
\ref{fig4}, where experimental measurements of $kA$ as function of
$\delta$ are shown for several chirp rates. As in Fig. \ref{fig3},
the distinction between phase locking at low values of $\mu$ and
phase unlocked states at high values of $\mu$ is clear. A closer
look at Fig. \ref{fig4}(a), however, reveals that the precise value
of $\mu_{cr}$ is undetermined. Two different runs with the identical
chirp rate of $\mu=0.27$ sec$^{-2}$ have qualitatively different
behavior: one decays, while the other phase locks into PAR. The
difference in these two runs stems from the dependence of $\mu_{cr}$
on the wave's initial amplitude $A_0$. As the waves evolve from
noise, we do not have experimental control over $A_0$. In Fig.
\ref{fig4}(b) we present fits of the initial stages of both of the
$\mu=0.27$ sec$^{-2}$ trajectories shown in Fig. \ref{fig4}(a) to
Eq. (\ref{passage}), where the sole fitting parameter is the value
of $A_0$. In both the phase-locked (${k}{A_0}=0.00269 \pm 0.00002$)
and unlocked (${k}{A_0}=0.0009 \pm 0.00004$) runs the experimental
points are indistinguishable from the theoretical predictions. Thus,
a difference of about $10\,\mu m$ in $A_0$ ($\sim 0.1\%$ of the
final, phase-locked amplitude) is sufficient to determine the wave's
long-time dynamics.
Similarly, we obtained $A_0$ for all of the runs presented in Fig.
\ref{fig4}(a) and, using these values, present the computed
functions ${k}{A}$ versus $\delta$ in Fig. \ref{fig4}(c), for the
parameter values used in Fig. \ref{fig4}(a). As in Fig. \ref{fig3},
the theory, which uses no other free parameters, is a strikingly
good description of the measurements, especially for values of
${k}{A}\leq{0.15}$, where the weak nonlinearity condition ${k}{A}
\ll {1}$ is well satisfied.  This is demonstrated in Fig.
\ref{fig4}(d) where a close comparison of theory and experiment is
performed for the three runs where phase locking failed. In the runs
where ${k}{A} \leq{0.15}$ throughout the entire experiment, theory
and experiment are nearly indistinguishable. The agreement
deteriorates when ${k}{A}> 0.2$, when the system is no longer in the
weakly nonlinear regime.

\textit{In summary}, one can control nonlinear Faraday waves by
employing PAR. The PAR technique remains operational for moderate
dissipation and well beyond the weak nonlinearity. It would be
interesting to extend it to multi-mode regimes and to other examples
of nonlinear waves.

This work was supported by the Israel Science Foundation (grants No.
194/02 and 107/05). We thank G. Cohen for advice on all aspects of
experiment.


\begin{thebibliography}{111}
\bibitem{pure} In our experiments the surface tension correction to the linear
wave frequency of the primary mode was less than 1\% for the whole
range of driving frequencies.
\bibitem {eps} The precise form of the $\varepsilon(t)$-dependence is of no importance,
as long as $\varepsilon(t)$ stays above the hysteretic lower tongue
and changes sufficiently slowly.
\bibitem{direct} The numerical results in Fig. \ref{fig3} (b) and (d) are within $15 \%$ from those
obtained by directly integrating  Eq. (\ref{motionchirp}).
\end{thebibliography}
\end{document}